\documentclass[aps,pra,showpacs,superscriptaddress,floatfix]{revtex4}
\usepackage{eurosym}
\usepackage{amssymb}
\usepackage{graphicx}

\begin{document}

\title{Non-equilibrium transition from dissipative quantum walk to classical
random walk}
\author{Marco Nizama and\ Manuel O. C\'{a}ceres}
\email{caceres@cab.cnea.gov.ar}
\affiliation{Centro At\'omico Bariloche, Instituto Balseiro and CONICET, 8400 Bariloche,
Argentina}
\date{\today }

\begin{abstract}
\ \newline

We have investigated the time-evolution of a free particle in interaction
with a phonon thermal bath, using the tight-binding approach. A dissipative
quantum walk can be defined and many important non-equilibrium decoherence
properties can be investigated analytically. The non-equilibrium statistics
of a pure initial state have been studied. Our theoretical results indicate
that the evolving wave-packet shows the suppression of Anderson's boundaries
(ballistic peaks) by the presence of dissipation. Many important relaxation
properties can be studied quantitatively, such as von Neumann's entropy and
quantum purity. In addition, we have studied Wigner's function. The
time-dependent behavior of the quantum entanglement between a free particle
-in the lattice- and the phonon bath has been characterized analytically.
This result strongly suggests the non-trivial time-dependence of the
off-diagonal elements of the reduced density matrix of the system. We have
established a connection between the quantum decoherence and the dissipative
parameter arising from interaction with the phonon bath. The time-dependent
behavior of quantum correlations has also been pointed out, showing
continuous transition from quantum random walk to classical random walk,
when dissipation increases.


\end{abstract}

\maketitle


\section{Introduction}

The nee to understand the evolution of a wave-packet (for instance, one
particle in a one dimensional lattice) has been motivated by many problems
in solid state physics \cite{kittel}, quantum information \cite{nielsen},
quantum open systems \cite{Alicki,Weiss,QN,Kampen} and quantum optics \cite%
{QN,zahringer}, among others. In solid state physics in particular, the
spread of a wave-packet from a highly localized initial state in the
tight-binding lattice approximation \cite{kittel,ashcroft} has drawn the
attention of many authors \cite{schreiber,yue}. Interestingly, the concept
of Quantum Walk (QW), borrowed from classical statistics \cite%
{aharanov,vK95,kempe,katsanos}, has the same properties as a tight-binding
free particle \cite{katsanos,esposito}. Two kinds of QW are considered in
the literature: discrete-time quantum coined walk \cite%
{zahringer,schreiber,aharanov,kempe,konno,chandrashekar,Romanelli,broome,shapira,dur,joye,schmitz}
and continuous-time QW \cite%
{yue,katsanos,esposito,manuel-karina,blumen2,perets,peruzzo}. In the former
(proposed by Aharonov et al.\cite{aharanov}), a two-level state, the
so-called "coin", rules the unitary discrete-time evolution of a particle
moving in a lattice. On the other hand, the evolution of the particle in \
the continuous-time QW is determined by a tight-binding like Hamiltonian 
\cite{esposito,blumen2}. It is not difficult to see, by simple comparison,
that a mapping between the \textit{tight-binding} Hamiltonian and the QW
model can be established (see Appendix A).

There are some important differences between a Classical Random Walk (CRW)
and a QW and these differences are well known in the literature \cite%
{yue,schreiber,Kampen,libro}. The most important of these differences is the
fact that for a tight-binding free particle the deviation $\sigma (t)$ of
the wave-packet becomes linear in time $\sigma (t)\sim t$ (QW), in contrast
to the CRW result that becomes similar to $\sigma (t)\sim \sqrt{t}$. This
issue can be resolved straightforwardly, by analyzing, in the Heisenberg
picture, the time evolution of the second moment of the position of the free
particle \cite{manuel-karina}. QWs have been studied for possible
applications in quantum information and quantum computation algorithms \cite%
{kempe,nielsen}. Discrete-time evolution in random media has recently been
used to study entanglement and quantum correlations in order to understand
the role of noise and the mechanism of decoherence between the internal and
spatial degrees of freedom in a QW \cite{chandrashekar}.

The study of a QW subjected to different sources of decoherence is an active
topic that has been considered by several authors, in particular, due to
their interest in understanding Laser cooling experiments \cite{Bouchaud03},
modeling Blinking Statistics \cite{Margolin}, and carrying out quantum
simulations \cite{Romanelli}. Experimental study of quantum decoherence in
discrete-time QW using single photons in the space was performed in \cite%
{broome}. These authors considered pure dephasing as a decoherence mechanism
and they could explore the quantum to classical transition by means of
tunable decoherence. In other theoretical studies \cite%
{yue,shapira,dur,konno,joye}, the authors analyzed discrete-time and
continuous-time QW in a random environment, and they could also study the
quantum-classical transition. In \cite{kendon} the authors study QW with
decoherence by analyzing a non-unitary evolution in QW. Experimental
analysis of QW in a random environment led to the study of quantum optical
devices \cite{schreiber,perets,schmitz,zahringer,peruzzo}; interestingly,
these experiments also show the non-classical behavior in DQW.

In classical statistics and via the Central Limit Theorem, the Gaussian
distribution plays a fundamental role in all random walks with finite mean
square displacement per step \cite{Kampen,libro}. In quantum statistics,
however, this analysis is much more complex because quantum thermal average
must be taken using the reduced density matrix. On the other hand, in the
Markov approximation, the time-evolution for the reduced density matrix
requires a much more complex infinitesimal generator, and in fact this
generator has been studied extensively, for many years, and is nowadays
called the Kossakoski-Lindblad generator \cite{kossa,lindblad}. One of the
most interesting facts that distinguishes quantum mechanics from classical
mechanics is the coherent superposition of distinct physical states. Many of
the non-intuitive aspects of the quantum theory of matter can be traced to
the coherent superposition feature. Related to this issue, an interesting
question arises: How does coherent superposition operate in the presence of
dissipation? These subjects have been important issues of research since the
pioneer works of Feynman and Vernon \cite{Feynmann1,Feynmann2} and, Caldeira
and Leggett \cite{caldeira}, among others, see for example the references
cited in: \cite{plenio,blum,spohn,QN,Kampen}.

An important fact in modeling a realistic QW is the inclusion of the quantum
thermal bath $\mathcal{B}$ from the very beginning in order to get a
dissipative open system \cite{Kampen,Weiss,QN}. To emphasize this fact we
call this continuous-time model a Dissipative Quantum Walk (DQW), and this
may be a mechanism of decoherence in a QW. Nevertheless there are many other
mechanism of decoherence, see for example \cite{kendon}. To our knowledge
this pioneer problem was well posed in van Kampen's paper \cite{vK95}, but
many other related approaches have also been presented in the literature 
\cite{esposito,whitfield}. The propagation of photons in waveguide lattices
have been studied in recent years \cite{perets,peruzzo}, and they are
possible scenarios where our present results can be applied.

In the present paper we study an open system in the Markov approximation to
continuous-time \cite{manuel,manuel-karina}, i.e., a quantum mechanical
particle that moves along a lattice by hopping while interacting with a
thermal phonon bath $\mathcal{B}$. We have chosen an interaction Hamiltonian
with the bath in such a way that this interaction produces the hopping in
the tight-binding particle. Therefore, we highlight some of the issues of
interpretation of the coherent superposition by tackling a soluble hopping
model. The asymptotic long-time regime of the quantum probability, Wigner%
\'{}%
s function, quantum entropy, quantum purity, etc., are characterized as a
function of the dissipation. The long-time decoherent behavior is also
explained in terms of the present dissipative hopping model. This model
analytically reproduces the (non-equilibrium) continuous transition from DQW
to CRW when the diffusion coefficient goes to infinity (i.e.: when the
temperature of the bath $T\rightarrow \infty $).

\subsection{A tight-binding open model}

We have considered a free particle model (tight-binding approximation, for
example see \cite{ashcroft}) constrained to a one-dimensional regular and
infinity lattice (in one band side) in interaction with a thermal bath of
phonon $\mathcal{B}$. The total Hamiltonian for this problem can be written
in the form \cite{vK95}:%
\begin{equation}
H_{\mathcal{T}}=\left( E_{0}\mathbf{1}-\Omega \frac{a+a^{\dag }}{2}\right)
+\sum\limits_{\nu =1}^{2}V_{\nu }\otimes B_{\nu }+H_{\mathcal{B}}.
\label{Hamiltonian}
\end{equation}

The first term corresponds to the \textit{tight-binding} Hamiltonian $H_{S}$
where $a$ and $a^{\dag }$ are translational operators in the Wannier bases $%
\left\vert s\right\rangle $ (it is easy to write these operators in second
quantization, for more details see appendix \ref{ap1:sec-quan}) and $\mathbf{%
1}$ is the identity operator. The second term is the interaction Hamiltonian
and corresponds to a linear coupling between phonon operators $%
B_{1}=B_{2}^{\dag }=\sum_{k}v_{k}\mathcal{B}_{k}$ and system operators $%
V_{1}=V_{2}^{\dag }=\hbar \Gamma a$, here $\Gamma >0$ is the coupling
parameter. The third term is the phonon Hamiltonian written in terms of
boson operators $\sum\limits_{k}\hbar \omega _{k}\mathcal{B}_{k}^{^{\dag }}%
\mathcal{B}_{k}$ \cite{manuel-karina,manuel,jpa}. Here $E_{0}$ is the 
\textit{tight-binding} energy of site and $\Omega $ the associated next
neighbor hopping energy.

The Quantum Master Equation (QME) for our DQW model can be obtained by
eliminating the variables of the quantum thermal bath and assuming for the
initial condition of the total density matrix $\rho _{T}(0)=\rho (0)\otimes
\rho _{\mathcal{B}}^{eq}$, where $\rho _{\mathcal{B}}^{eq}$ is the
equilibrium density matrix of the quantum bath $\mathcal{B}$. Then, in the
Markov approximation, using $\left[ a,a^{\dag }\right] =0$ and $a^{\dag }a=%
\mathbf{1}$, the evolution equation for the reduced density matrix is \cite%
{manuel,manuel-karina}: 
\begin{equation}
\dot{\rho}\equiv \frac{d\rho }{dt}=\frac{-i}{\hbar }\left[ H_{eff},\rho %
\right] +D\left( a\rho a^{\dag }+a^{\dag }\rho a-2\rho \right) ,  \label{QME}
\end{equation}%
with a trivial effective Hamiltonian: $H_{eff}=H_{S}-\hbar \omega _{c}%
\mathbf{1}.$ The diffusion constant is given in terms of the quantum thermal
bath temperature and the coupling constant in the form: 
\[
D\propto \Gamma ^{2}k_{B}T/\hbar , 
\]%
the additive energy $\hbar \omega _{c}$ is related to the Caldeira and
Legett frequency cut-off in the Ohmic approximation \cite{caldeira}. For
simplicity we can add an additive constant to the \textit{tight-binding}
Hamiltonian $-E_{0}+\omega _{c}\hbar +\Omega $. This assumption does not
change the general results and finally we can write: 
\begin{equation}
H_{eff}=\Omega \left( \mathbf{1}-\frac{a+a^{\dag }}{2}\right) ,  \label{He}
\end{equation}%
as was presented in previous references \cite{vK95,manuel,manuel-karina}. It
can be noted from Eq.(\ref{QME}) that as $D\rightarrow 0$ ($T\rightarrow 0$%
), the von Neumann equation is recovered (unitary evolution).

\subsubsection{\protect\bigskip On the second moment of the DQW}

\label{subsec:mom}

From the QME Eq.(\ref{QME}), we can obtain the dynamics of any operator, in
particular here we are interested in the evolution of the dispersion of the
position operator $\mathbf{q}$, which in the Wannier basis has the matrix
elements ($s$ is an eigenvalue of $\mathbf{q}$):%
\begin{equation}
\left\langle s\right\vert \mathbf{q}\left\vert s^{\prime }\right\rangle =s\
\delta _{s,s^{\prime }},  \label{position}
\end{equation}%
note that $\mathbf{q}$ is defined as a dimensionless position operator with
lattice parameter $\epsilon =1$. Then the quantum thermal time-evolution of
the first and second moments can be calculated straightforwardly. In the
Heisenberg representation we get%
\begin{eqnarray}
\frac{d}{dt}\mathbf{q}(t) &=&\frac{-i}{\hbar }\left[ \mathbf{q},H_{eff}%
\right]  \label{1mom} \\
\frac{d}{dt}\mathbf{q}^{2}(t) &=&\frac{-i}{\hbar }\left[ \mathbf{q}%
^{2},H_{eff}\right] +2D\mathbf{1}.  \label{2mom}
\end{eqnarray}%
Where we have used that $\left[ \mathbf{q},H_{eff}\right] =\frac{\Omega }{2}%
\left( a-a^{\dag }\right) $ and that $a(t)=a(0),a^{\dag }(t)=a^{\dag }(0)$,
then we get for the time evolution of the position operator:%
\[
\mathbf{q}(t)=\frac{-i\Omega }{2\hbar }\left( a-a^{\dag }\right) t+\mathbf{q}%
(0). 
\]

It is simple to realize, just from a physical point of view, why there is no
dissipative term in the equation of motion for $\mathbf{q}(t)$. Taking the
thermal average in Eq. (\ref{1mom}) we get $\frac{d}{dt}\left\langle \mathbf{%
q}(t)\right\rangle =\frac{\Omega }{\hbar }\sum_{s=-\infty }^{s=\infty }%
\mbox{Im}\lbrack \rho _{s,s-1}(t)],$ then introducing the explicit solution
for the density matrix, Eq. (\ref{rhog}), and using the properties of the
Bessel function (with integer indices) $\sum_{s=-\infty }^{s=\infty
}J_{s+n}\left( x\right) J_{s+m}\left( x\right) =\delta _{n,m}$. We get that $%
\frac{d}{dt}\left\langle \mathbf{q}(t)\right\rangle =0$ indicating the
conservation of reflection symmetry at the localized initial condition,
otherwise if $\frac{d}{dt}\left\langle \mathbf{q}(t)\right\rangle \neq 0,$
this would destroy the reflection symmetry principle.

The quantum thermal statistical average -of any operator- in the Heisenberg
picture can be written as $\left\langle \mathbf{A}(t)\right\rangle =\mbox{Tr}%
\left[ \mathbf{A}(t)\rho (0)\right] $. Then for the variance of the DQW we
get: 
\begin{equation}
\sigma (t)^{2}=\left\langle \mathbf{q}(t)^{2}\right\rangle -\left\langle 
\mathbf{q}(t)\right\rangle ^{2}=\frac{1}{2}\left( \frac{\Omega t}{\hbar }%
\right) ^{2}+2Dt,  \label{variance}
\end{equation}%
which is the expected dissipative result \cite%
{vK95,manuel-karina,manuel,Kampen}. From the Eqs. (\ref{1mom}) and (\ref%
{2mom}), it is possible to see that von Neumann's term gives a contribution
of the form $\propto t^{2}$ for the time-evolution of the second moment,
this is a well known quantum result. In fact for the null dissipation case,\ 
$D=0,$\ we get that Anderson's boundaries (ballistic peak) movement is
controlled by the linear law of the deviation of the wave-packet:%
\[
\sigma (t)=\sqrt{\left\langle \mathbf{q}(t)^{2}\right\rangle -\left\langle 
\mathbf{q}(t)\right\rangle ^{2}}=\frac{1}{\sqrt{2}}\frac{\Omega t}{\hbar },\ 
\text{if }D=0. 
\]%
Therefore, we can associate the quantity $V_{A}=\frac{1}{\sqrt{2}}\frac{%
\Omega }{\hbar }$ with the velocity of Anderson's boundaries in a
one-dimensional regular lattice.

In the appendix \ref{ap2:momen} we have calculated the second moment of the
position operator by using the characteristic function, which is useful for
calculating all quantum thermal moments.

In the next section we will present, in detail, results concerning the
probability profile of our quantum open model ($D\neq 0$), i.e., \textit{a
dissipative tight-binding free particle}. We will study the time-evolution
of the reduced density matrix \cite{jpa,ukranian}, characterize its
decoherence, and solve, analytically, some correlation functions associated
with the coherent superposition feature.

\section{Time evolution of the DQW}

\subsection{General properties}

In order to consider the evolution of our free particle in interaction with
the quantum thermal bath $\mathcal{B}$, we have to solve the QME, Eq.(\ref%
{QME}), for any time. This can be done in the Fourier representation. Let
the Fourier "\textit{bra-ket}" be defined in terms of the Wannier "\textit{%
bra-ket}" in the form: 
\begin{eqnarray*}
\left\vert k\right\rangle &=&\frac{1}{\sqrt{2\pi }}\sum\limits_{s=-\infty
}^{\infty }e^{iks}\left\vert s\right\rangle , \\
\left\langle k\right\vert &=&\frac{1}{\sqrt{2\pi }}\sum\limits_{s=-\infty
}^{\infty }e^{-iks}\left\langle s\right\vert .
\end{eqnarray*}%
Then the QME adopts an explicit form:%
\begin{equation}
\left\langle k_{1}\right\vert \!\dot{\rho}\!\left\vert k_{2}\right\rangle
\!\!=\!\!\left[ \!\frac{-i}{\hbar }\!\left( \mathcal{E}_{k_{1}}\!-\!\mathcal{%
E}_{k_{2}}\right) \!+\!2D\!\left( \cos \left( k_{1}\!-\!k_{2}\right)
\!-\!1\right) \!\right] \!\!\left\langle k_{1}\right\vert \!\rho
\!\left\vert k_{2}\right\rangle ,  \label{QMEk}
\end{equation}%
where $\mathcal{E}_{k}=\Omega \left\{ 1-\cos k\right\} .$ Note that the
diagonal elements $\rho _{k,k}(t)\equiv \langle k\arrowvert\rho (t)%
\arrowvert
k\rangle $ are constant in time, $\dot{\rho}_{k,k}(t)=0$, for example, for a
localized initial condition $\rho (0)=\arrowvert s_{0}\rangle \langle s_{0}%
\arrowvert$ (with $s_{0}=0$) we get: 
\begin{equation}
\rho _{k,k}(t)=\rho _{k,k}(0)=\frac{1}{2\pi },\quad \forall t,
\label{probgk}
\end{equation}%
even so, we can define a pseudo-momentum operator $\mbox{ \bf p}\equiv \frac{%
m}{i\hbar }[q,H_{s}]$, where $m$ represents the mass of the free particle in
the model. We can calculate any moment of the pseudo-momentum operator, $%
\langle \mbox{\bf p}^{j}\rangle =\mbox{Tr[\bf p}^{j}\rho (t)]$ for $%
j=1,2,\cdots $. For the case $j=1$, we obtain $\langle \mbox{\bf p}\rangle
=0 $ and for $j=2$, we get $\langle \mbox{\bf p}^{2}\rangle =\frac{1}{2}(%
\frac{\Omega }{\hbar })^{2}m^{2}$. Then we can define the quantum thermal
second moment of the velocity $\mathbf{v}$ in the following way: $\langle 
\mbox{\bf
v}^{2}\rangle =\frac{1}{2}(\frac{\Omega }{\hbar })^{2}$ and so we re-obtain
the velocity of Anderson's boundaries $V_{A}=\frac{1}{\sqrt{2}}\frac{\Omega 
}{\hbar }$ (see subsection "On the second moment of the DQW").

To solve Eq.(\ref{QMEk}) we define the function: 
\begin{equation}
\mathcal{F}\!\left( k_{1},k_{2}\right) =i\frac{\Omega }{\hbar }\left( \cos
(k_{1})\!-\!\cos (k_{2})\right) +2D\!\left( \cos
(k_{1}\!-\!k_{2})\!-\!1\right) ,  \label{Fk1k2}
\end{equation}%
the general solution of the QME\ can be written as:%
\begin{equation}
\left\langle k_{1}\right\vert \rho (t)\left\vert k_{2}\right\rangle =\rho
(0)_{k_{1}k_{2}}\ \exp \left( \mathcal{F}\left( k_{1},k_{2}\right) t\right) .
\label{evo2}
\end{equation}

In order to study the suppression of Anderson's boundaries by the presence
of dissipation, it is convenient to go back to the Wannier representation ($%
\arrowvert s\rangle =\frac{1}{\sqrt{2\pi }}\int_{-\pi }^{\pi }dk\ e^{-iks}%
\arrowvert k\rangle $). Adopting Eq.(\ref{probgk}), as initial condition for
the density matrix, we get%
\begin{eqnarray}
\left\langle s_{1}\right\vert \!\rho (t)\!\left\vert s_{2}\right\rangle &=&%
\frac{1}{2\pi }\int_{-\pi }^{\pi }dk_{1}\int_{-\pi }^{\pi
}dk_{2}e^{i(k_{1}s_{1}-k_{2}s_{2})}\left\langle k_{1}\right\vert \!\rho
(t)\!\left\vert k_{2}\right\rangle  \nonumber  \label{roL1L2} \\
\!\! &=&\!\!\left( \!\frac{1}{2\pi }\!\right) ^{\!\!2}\!\!\!\!\int_{-\pi
}^{\pi }\!\!\!\!dk_{1}\!\!\int_{-\pi }^{\pi }\!\!\!\!dk_{2}e^{\left[ i\left(
k_{1}-k_{2}\right) (s_{1}-s_{2})\right] }e^{\mathcal{F}\left(
k_{1},k_{2}\right) t}.  \nonumber \\
&&
\end{eqnarray}

We can solve Eq.(\ref{roL1L2}) analytically if we consider Bessel's function
properties: $e^{iz\cos \theta }=\sum_{n=-\infty }^{\infty
}i^{n}J_{n}(z)e^{in\theta }$; $e^{z\cos \theta }=\sum_{n=-\infty }^{\infty
}I_{n}(z)e^{in\theta }$, where $J_{n}$ and $I_{n}$ are Bessel functions of
integer order. Using the following relations $J_{-n}(x)=(-1)^{n}J_{n}(x)$, $%
J_{n}(-x)=(-1)^{n}J_{n}(x)$ and $I_{-n}(x)=I_{n}(x)$, $%
I_{n}(-x)=(-1)^{n}I_{n}(x)$, where $n$ is an integer \cite%
{abramowitz,evangelidis}, we have obtained an analytical expression for $%
\langle s_{1}|\rho (t)|s_{2}\rangle $ (for more details, see appendix \ref%
{ap:matrix}):

\begin{eqnarray}
\langle s_{1}|\rho (t)|s_{2}\rangle \!\!
&=&\!i^{(s_{1}-s_{2})}e^{-2Dt}\!\!\!\!\sum_{n=-\infty }^{\infty
}\!\!\!\!J_{s_{1}+n}\!\!\left( \!\frac{\Omega t}{\hbar }\!\right)
\!\!J_{s_{2}+n}\!\!\left( \!\frac{\Omega t}{\hbar }\!\right)
\!\!I_{n}\!\left( 2Dt\right) .  \label{rhog} \\
&&  \nonumber
\end{eqnarray}

Using $\sum_{n=-\infty }^{\infty }J_{n}^{2}(x)=1$ and $\sum_{-\infty
}^{\infty }I_{n}(x)=e^{x}$ \cite{abramowitz}, we can check that Eq.(\ref%
{rhog}) fulfills the normalization condition $\mbox{Tr}\lbrack \rho
(t)]=\sum_{s=-\infty }^{\infty }\langle s|\rho (t)|s\rangle =1$, $\forall D$%
, and the probability of finding the particle in the site $s$ in the lattice
is: 
\begin{eqnarray}
P_{s}(t) &\equiv &\langle s|\rho (t)|s\rangle =e^{-2Dt}\!\!\sum_{n=-\infty
}^{\infty }\left[ J_{s+n}\!\!\left( \!\frac{\Omega t}{\hbar }\!\right) %
\right] ^{2}I_{n}\!\left( 2Dt\right) ,\;DQW.  \label{probg} \\
&&  \nonumber
\end{eqnarray}%
It is straightforward to note that Eq.(\ref{rhog}) contains all the
information concerning the transition from DQW to CRW. In fact this
transition is a genuine non-equilibrium one because any behavior
characterizing the quantum to classical transition will be given in terms of
the time-evolution of the density matrix, which of course is not a Gibbsian
density matrix \cite{Weiss,Alicki,QN}.

If $D=0$ (without dissipation, i.e., a closed system), we recover the QW and
in this case the matrix elements $\langle s_{1}|\rho (t)|s_{2}\rangle $
reduce to the form (using $I_{n}(0)=\delta _{n,0}$, where $n$ is an integer):

\begin{eqnarray}
\left\langle s_{1}\right\vert \rho (t)\left\vert s_{2}\right\rangle _{D=0}\!
&=&i^{(s_{1}-s_{2})}J_{s_{1}}\!\left( \!\frac{\Omega t}{\hbar }\!\right)
\!J_{s_{2}}\!\left( \!\frac{\Omega t}{\hbar }\!\right) ,\;QW.  \label{rhoD_0}
\\
&&  \nonumber
\end{eqnarray}

\bigskip

In the case $D\rightarrow \infty $, we can re-obtain exactly the classical
probability \cite{libro,Kampen}, where the off-diagonal elements of $\rho
(t) $ in Wannier basis are equal to zero. Alternatively, consider the limits 
$\Omega \rightarrow 0$ and $\hbar \rightarrow 0$ in such a way that:%
\[
\lim_{\Omega \rightarrow 0,\hbar \rightarrow 0}\frac{\Omega }{\hbar }%
\rightarrow 0, 
\]%
then from Eq.(\ref{probg}) and using $J_{s+n}(0)=\delta _{s+n,0}$, where $%
s+n $ is an integer it follows (for any finite time $t$) that

\[
\lim_{\Omega \rightarrow 0,\hbar \rightarrow 0}\langle s|\rho (t)|s\rangle
\rightarrow e^{-2Dt}\!\!\sum_{n=-\infty }^{\infty }\left[ \delta _{s,-n}%
\right] ^{2}I_{n}\!\left( 2Dt\right) =e^{-2Dt}\!\!\ I_{s}\!\left( 2Dt\right)
, 
\]%
which is just the probability of the CRW, i.e.,:%
\begin{equation}
P_{s}(t)=e^{-2Dt}\ I_{s}\left( 2Dt\right) ,\;CRW.  \label{rhoD_infty}
\end{equation}%
We note from Eq.(\ref{rhoD_infty}), when $t\rightarrow \infty $ that we
re-obtain the well known Gaussian asymptotic scaling for the CRW probability 
$P_{s}(t\rightarrow \infty )\rightarrow \frac{t^{-1/2}}{\sqrt{4\pi D}}$ (we
have used the asymptotic limit of $I_{n}(x)\approx \frac{e^{x}}{\sqrt{2\pi x}%
}$, for $x\rightarrow \infty $ \cite{abramowitz}).

Here we have considered it appropriate to define a new parameter such that $%
r_{D}=\frac{2D}{\Omega /\hbar }$ (rate of characteristic energy scales in
the system) and $t^{\prime }=\frac{\Omega }{\hbar }t$ a dimensionless time,
in order to plot the analytical expression (\ref{probg}). In Fig.\ref%
{fig-prob2D} we show the probability at the site $s$ $P_{s}(t)=\langle s%
\arrowvert\rho (t)\arrowvert s\rangle $, for different values of the
dissipative parameter $r_{D}$ (see, Eq.(\ref{probg})). Similar plots for the
probability profile in the presence of dissipation have been analyzed by
Esposito et al. \cite{esposito}.

\begin{figure}[t]
\includegraphics[width=0.95 \columnwidth,clip]{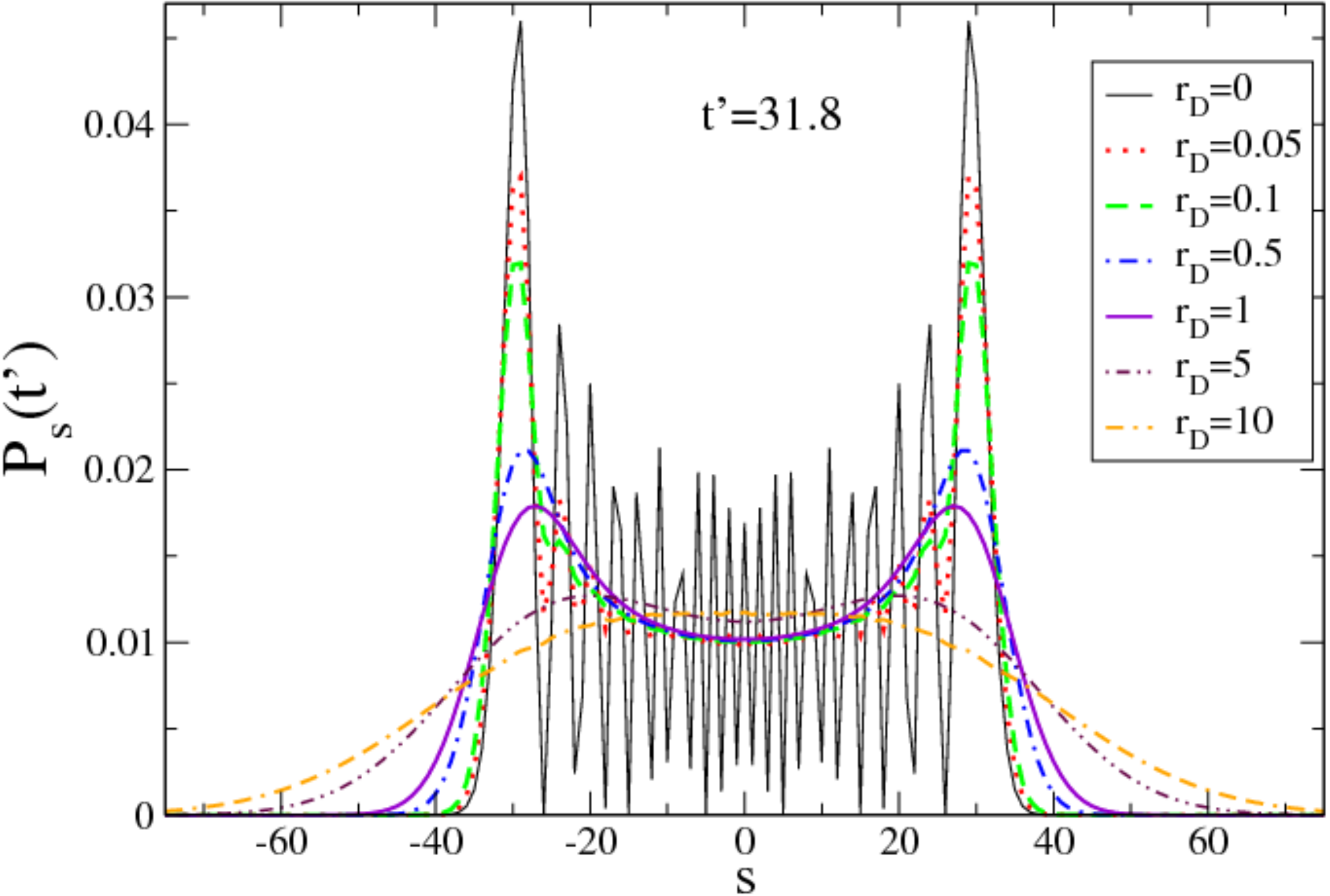}
\caption{Probability $P_{s}(t^{\prime })$ for an initial condition $\protect%
\rho (0)=\arrowvert s_{0}\rangle \langle s_{0}\arrowvert$ (where $s_{0}=0$)
as\ a function of position $s$, for $r_{D}=0$, $0.1$, $0.5$, $1$, $5$, $10$
and $t^{\prime }=31.8$, where we have defined the new variables $r_{D}=\frac{%
2D}{\Omega /\hbar }$ and $t^{\prime }=\frac{\Omega }{\hbar }t$
(dimensionless time). }
\label{fig-prob2D}
\end{figure}

Note from Fig.\ref{fig-prob2D} that for $D=0$ ($r_{D}=0$), the system is not
interacting with the phonon bath; in this case, we get a closed system, and
therefore the evolution of the wave-packet is not diffusive but is
ballistic, Eq.(\ref{rhoD_0}). So for $t^{\prime }=31.8$ we observe two
maximum peaks in $s_{p}\approx \pm 29$, which correspond to ballistic peaks
(Anderson's boundaries), and far away from these peaks for $\arrowvert s%
\arrowvert>s_{p}$ the probability quickly goes to zero. In the case that $%
\arrowvert s\arrowvert<s_{p}$, we easily see oscillatory behavior because of
the quantum behavior of the system. If the dissipative term is small $%
r_{D}<<1$ ($r_{D}=0.05$) the oscillatory behavior starts to decrease and
when the dissipation is of the order of the hopping energy ($r_{D}=1$) or
larger ($r_{D}=5,10$, see Fig.1), the dissipation dominates in the system
and the quantum character vanishes. In this case the wave-packet tends to a
Gaussian form. This is the regimen for the CRW, Eq.(\ref{rhoD_infty}).

In the remainder of the paper we are going to study the probability profile
of the DQW, the quantum purity, the Wigner function, and von Newmann's
entropy as a function of the dissipative parameter $r_{D}$. In the Wigner
section we are also going to introduce a criterion to describe the
quantum-classical transition (see inset of figure 5). In a future work we
will present analytical and numerical results such as the concurrence,
negativity, etc., as a function of $r_{D}$ in order to study the decoherence
and the entanglement in a bipartite system related to our DQW.

\begin{figure}[t]
\includegraphics[height=20cm,width=8.5cm]{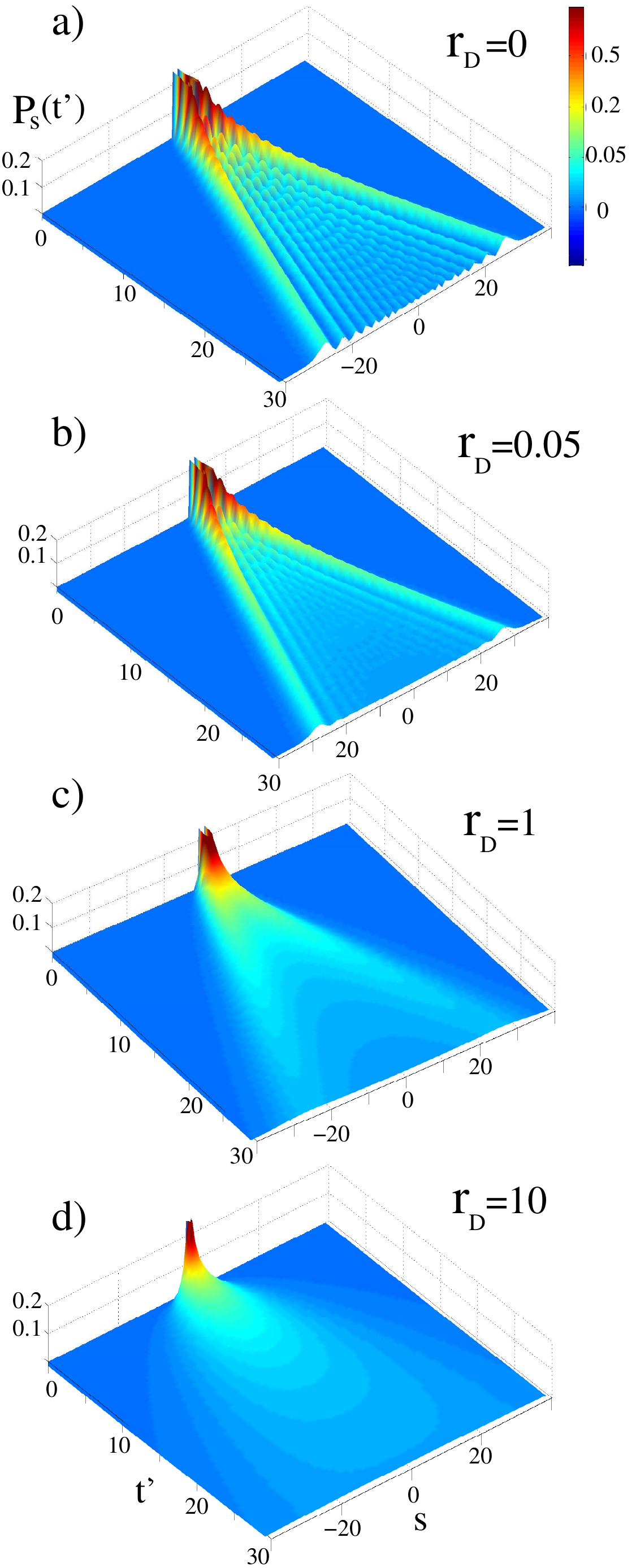}
\caption{(Color online) Representation in 3D of probability $P_{s}(t^{\prime
})$ for an initial condition $\protect\rho (0)=\arrowvert s_{0}\rangle
\langle s_{0}\arrowvert$ (where $s_{0}=0$) as a function of position $s$ and 
$t^{\prime }$, for $r_{D}=0$, $0.05$, $1$, $10$ [(a)-(d)], where the blue
regions indicate, approximately, value zero and red regions have high value
for probability. The variables $r_{D}$ and $t^{\prime }$ are the same as in
Fig.\protect\ref{fig-prob2D}. }
\label{fig-prob3D}
\end{figure}

In Fig.\ref{fig-prob3D} we show the same\ facts as in Fig.\ref{fig-prob2D}
(for some values of $r_{D}$) but in three dimensions (3D); i.e., we have
included an extra axis for the time $t^{\prime }$ (this kind of graphic
representation is usually called quantum carpet). We cut the axis of the
probability for convenience (for example we don't show the probability for $%
t^{\prime }=0,$ $P_{s}(0)=1$). In this quantum carpet we observe the
transition from quantum regimen (Fig.\ref{fig-prob2D} (a)) to classical
regimen (Fig.\ref{fig-prob2D} (d)). Oscillatory behavior for small values of 
$D$ ($r_{D}<<1$) is observed, and the oscillations in the probability start
to disappear when $r_{D}$ is larger than one.

On the other hand, the quantum purity \cite{nielsen} ($\mathcal{P}%
_{Q}(t)\equiv \mbox{Tr}\lbrack \rho (t)^{2}]$), a quantity that provides
information about whether the state is pure ($\mathcal{P}_{Q}=1$) or mixed ($%
\mathcal{P}_{Q}<1$), can be calculated analytically. In the present model
quantum purity can be calculated using Bessel properties \cite%
{abramowitz,evangelidis,martin}:

\begin{eqnarray}
\mathcal{P}_{Q}(t) &=&\mbox{Tr}\lbrack \rho (t)^{2}]=\sum_{s_{1}=-\infty
}^{\infty }\sum_{s_{3}=-\infty }^{\infty }\langle s_{1}|\rho
(t)|s_{3}\rangle \langle s_{3}|\rho (t)|s_{1}\rangle  \nonumber \\
&=&i^{(s_{1}-s_{3})}e^{-2Dt}\!\!\!\!\sum_{n=-\infty }^{\infty
}\!\!\!\!J_{s_{1}+n}\!\!\left( \!\frac{\Omega t}{\hbar }\!\right)
\!\!J_{s_{3}+n}\!\!\left( \!\frac{\Omega t}{\hbar }\!\right)
\!\!I_{n}\!\left( 2Dt\right)  \nonumber \\
&&\times i^{(s_{3}-s_{1})}e^{-2Dt}\!\!\!\!\sum_{n^{\prime }=-\infty
}^{\infty }\!\!\!\!J_{s_{3}+n^{\prime }}\!\!\left( \!\frac{\Omega t}{\hbar }%
\!\right) \!\!J_{s_{1}+n^{\prime }}\!\!\left( \!\frac{\Omega t}{\hbar }%
\!\right) \!\!I_{n^{\prime }}\!\left( 2Dt\right) .  \nonumber
\end{eqnarray}

Using the relations $\sum_{n=-\infty }^{\infty }J_{n+m}(x)J_{n+m^{\prime
}}(x)=\delta _{m,m^{\prime }}$ and $\sum_{n=-\infty }^{\infty
}I_{n}^{2}(x)=I_{0}(2x)$, we get 
\[
\mathcal{P}_{Q}(t)=e^{-4Dt}I_{0}\left( 4Dt\right) . 
\]

The quantum purity takes the value one for $D=0$ (without dissipation) for
all time, and for the case $D\neq 0$ the quantum purity takes values smaller
than one for $t>0$ (mixed state). 
For $D\neq 0$, the quantum purity takes on the asymptotic power law behavior 
$\mathcal{P}_{Q}(t)\sim t^{-1/2}$, for $t\rightarrow \infty $ (using $%
I_{n}(x)\approx \frac{e^{x}}{\sqrt{2\pi x}}$, for $x\rightarrow \infty $ 
\cite{abramowitz}), in agreement with previous results presented in the
references \cite{manuel}.

Another important measure that will indicate the presence of quantum
behavior is the Wigner function. In the next subsection, we will study this
function.

\subsection{The Wigner function}

The Wigner function was originally formulated as a quasiprobability for the
position and momentum of the particle in quantum mechanics. For continuous
variables $X$ and $K$, representing position and momentum in the phase
space, the Wigner function is given by \cite{wigner,hillery-wigner}: 
\begin{equation}
W(X,K,t)=\frac{1}{\pi }\int dY\langle X+Y\arrowvert\rho (t)\arrowvert %
X-Y\rangle e^{2iKY},  \label{eq-wig}
\end{equation}%
where $\rho (t)$ is the time-dependent density operator. The Wigner function
could have negative values whereby it is considered a quasi joint
probability of $X$ and $K$, and the marginal distribution for $X$ and $K$
can be obtained in the usual form: 
\begin{eqnarray}
\int dKW(X,K,t) &=&\langle X\arrowvert\rho (t)\arrowvert X\rangle
,\;\;\;\;\;\;\mbox{(a)}  \nonumber \\
\int dXW(X,K,t) &=&\langle K\arrowvert\rho (t)\arrowvert K\rangle
,\;\;\;\;\;\;\mbox{(b)}  \label{eq-marg}
\end{eqnarray}%
where $\langle X\arrowvert\rho (t)\arrowvert X\rangle $ and $\langle K%
\arrowvert\rho (t)\arrowvert K\rangle $ are the time-dependent probability
densities for $X$ and $K$. The normalization condition for $\rho (t)$ can be
checked from Eq.(\ref{eq-wig}): 
\begin{equation}
\int dX\int dKW(X,K,t)=\mbox{Tr}\lbrack \rho (t)]=1.  \label{eq-wig-nor}
\end{equation}

In our DQW model the space is an infinite one-dimensional regular lattice.
Thus, in this case, we propose the Wigner function as: 
\begin{equation}
W(s,k,t)=\frac{1}{2\pi }\sum_{s^{\prime }=-\infty }^{\infty }\langle
s+s^{\prime }\arrowvert\rho (t)\arrowvert s-s^{\prime }\rangle
e^{iks^{\prime }},  \label{eq-wig-dis}
\end{equation}%
where $\arrowvert s\rangle $ is a Wannier basis. This definition fulfills
the required conditions for the Wigner function, see Eqs. (\ref{eq-marg})
and (\ref{eq-wig-nor}), using $\sum_{s}$ instead of $\int dX$, where $k\in
\lbrack -\pi ,\pi ]$ (first Brillouin zone).

Using Eq.(\ref{rhog}) in Eq.(\ref{eq-wig-dis}), we can write the Wigner
function as follows: 
\begin{equation}
W(s,k,t)\!=\!\frac{e^{-2Dt}}{2\pi }\!\!\!\!\sum_{n,s^{\prime }=-\infty
}^{\infty }\!\!\!\!i^{2s^{\prime }}e^{iks^{\prime }}J_{s+s^{\prime
}+n}\!\!\left( \!\frac{\Omega t}{\hbar }\!\right) \!\!J_{s-s^{\prime
}+n}\!\!\left( \!\frac{\Omega t}{\hbar }\!\right) \!\!I_{n}\!\left(
2Dt\right) ,  \label{eq-wig-dis0}
\end{equation}%
then, after some algebra and using Bessel's properties \cite%
{abramowitz,evangelidis,martin} ($\sum_{n=-\infty }^{\infty }e^{in\gamma
}J_{n+m}(x)J_{n}(x)=J_{m}(2x\sin (\gamma /2))e^{i\beta m}$, where $\beta
=\pi /2-\gamma /2$), we find:

\begin{equation}
W(s,k,t)=\frac{e^{-2D t}}{2 \pi} \!\!\!\! \sum_{n=-\infty}^{\infty} \!\!
\!\! J_{2s + 2 n } \!\! \left(2 \! \frac{\Omega t}{\hbar } \sin \frac{k}{2}%
\! \right) \!\! I_{n }\! \left( 2 D t \right).  \label{eq-wig-dis1}
\end{equation}

Using Eqs.(\ref{eq-marg}-a) and (\ref{eq-marg}-b) in Eq.(\ref{eq-wig-dis1}),
we recover the probability densities for position $s$ and momentum $k$
respectively (see, eqs. (\ref{probg}) and (\ref{probgk})).

Interestingly, the Wigner function has information about the transition from
DQW to CRW. Analyzing the case $D=0$ (without dissipation in the system) the
Wigner function adopts the following form: 
\begin{equation}
W(s,k,t)_{QW}=\frac{1}{2\pi }J_{2s}\!\!\left( 2\!\frac{\Omega t}{\hbar }\sin 
\frac{k}{2}\!\right) ,\;\;\;\;D=0.  \label{eq-wig-qua}
\end{equation}

In the pure diffusive regimen ($\Omega =0$, i.e., the CRW), the Wigner
function can be written as: 
\begin{equation}
W(s,k,t)_{CRW}=\frac{e^{-2Dt}}{2\pi }I_{s}\!\left( 2Dt\right)
,\;\;\;\;\Omega =0.  \label{eq-wig-clas}
\end{equation}

Then from Eqs.(\ref{eq-wig-qua}) and (\ref{eq-wig-clas}), the Wigner
function can be re-written as: 
\begin{equation}
W(s,k,t)=2\pi \sum_{n=-\infty }^{\infty }W(s+n,k,t)_{QW}\ W(n,k,t)_{CRW}, 
\nonumber
\end{equation}%
then, changing $n\rightarrow -n$ and considering that $I_{-n}(x)=I_{n}(x)$,
we get 
\begin{equation}
W(s,k,t)=2\pi \sum_{n=-\infty }^{\infty }W(s-n,k,t)_{QW}\ W(n,k,t)_{CRW},
\label{eq-wig-gen}
\end{equation}%
this expression shows that the Wigner function of the DQW is given in terms
of a non-trivial space convolution operation between the QW ($D=0$) and the
CRW ($\Omega =0$). For the case $D\neq 0$ this expression shows the
nonlocality of the quantum mechanics of a free particle interacting with the
quantum thermal $\mathcal{B}$.

\begin{figure}[t]
\includegraphics[height=8.5cm,width=8.5cm]{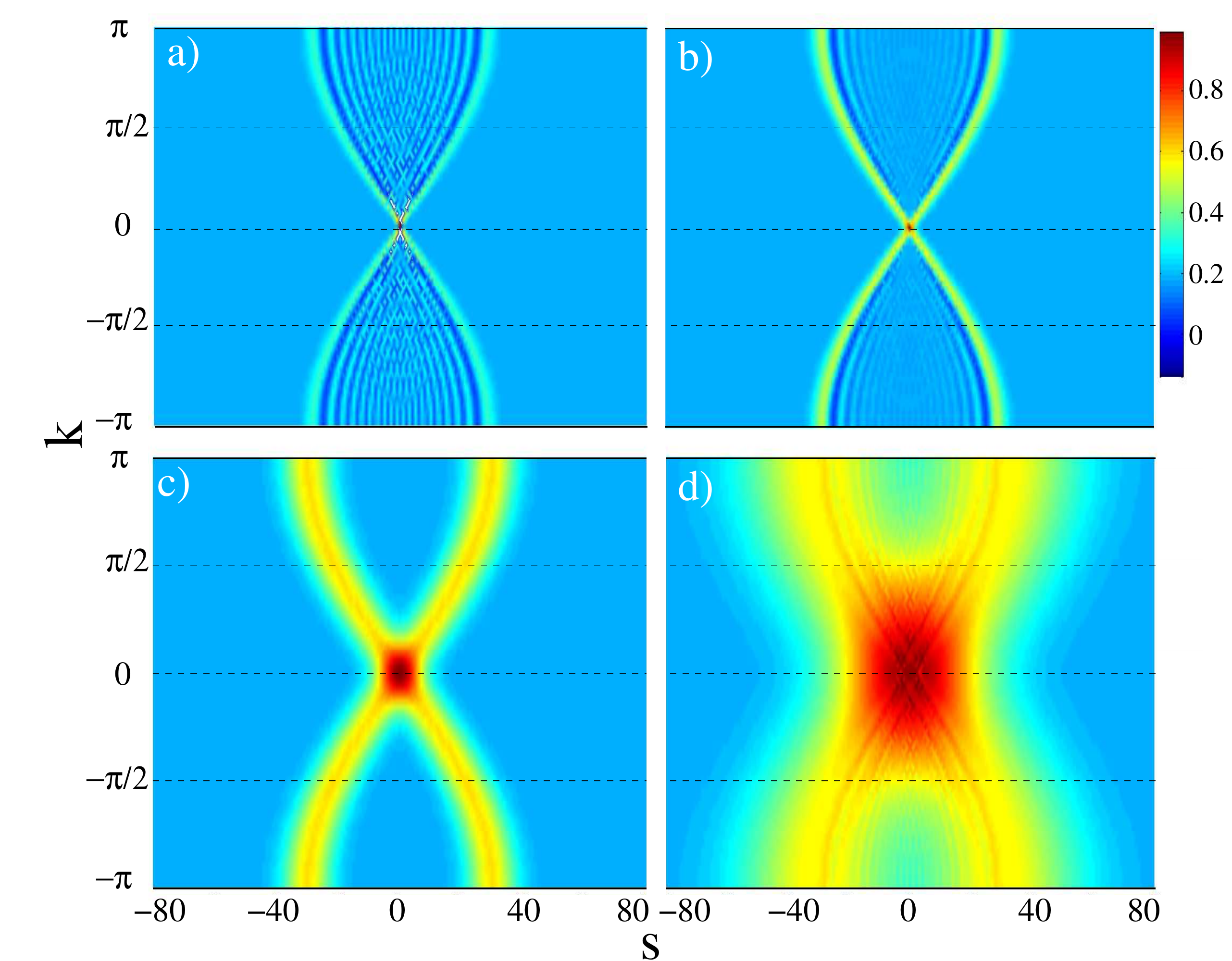}
\caption{ (Color online) Normalized Wigner function $W(s,k,t^{\prime
})/W_{max}$ as a function of position $s$ and momentum $k$, for $r_{D}=0$, $%
0.05$, $1$, $10$ ([(a) $-$ (d)]) and $t^{\prime }=30$, where $W_{max}$ is
the maximum value of Wigner function for each value of $r_{D}$. We have
considered $r_{D}=\frac{2D}{\Omega /\hbar }$ and $t^{\prime }=\frac{\Omega }{%
\hbar }t$ (dimensionless). The regions in blue indicate negative values, the
ones in sky blue indicate, approximately, zero value and the red regions
have the highest values of the Wigner function. }
\label{fig-wig}
\end{figure}

In Fig.\ref{fig-wig}, we show the Wigner function from Eq.(\ref{eq-wig-dis1}%
) or Eq.(\ref{eq-wig-gen}). We have normalized the Wigner function with
respect to its maximum value $W_{max}$ for each value of $r_{D}$ (i.e.: we
use $W(s,k,t^{\prime })/W_{max}$, where $t^{\prime }=\frac{\Omega }{\hbar }t$
and $r_{D}=\frac{2D}{\Omega /\hbar }$). Similar Wigner's quantum carpets
have been analyzed for a QW on a ring without dissipation in \cite{blumen3}.
In Fig.\ref{fig-wig}(a), we observe the quantum behavior for $r_{D}=0$. In
this case the Wigner function has negative and positive values, and it shows
remarkable oscillatory behavior (see Eq.(\ref{eq-wig-qua})). In the case of
small values of the dissipation, $r_{D}<<1$, we observe (Fig.\ref{fig-wig}%
(b), with $r_{D}=0.05<<1$) that the oscillatory behavior starts to diminish
and for values of $r_{D}\gtrsim 1$ (fig \ref{fig-wig}(c)-(d)) the
dissipative term is dominant. In this case, the Wigner function takes only
positive values, therefore it is a well-defined joint probability, and the
system resembles a CRW when $r_{D}$ tends to infinity. In this case, the
system is diffusive (see Eq.(\ref{eq-wig-clas})).

\begin{figure}[t]
\includegraphics[height=8.5cm,width=8.5cm]{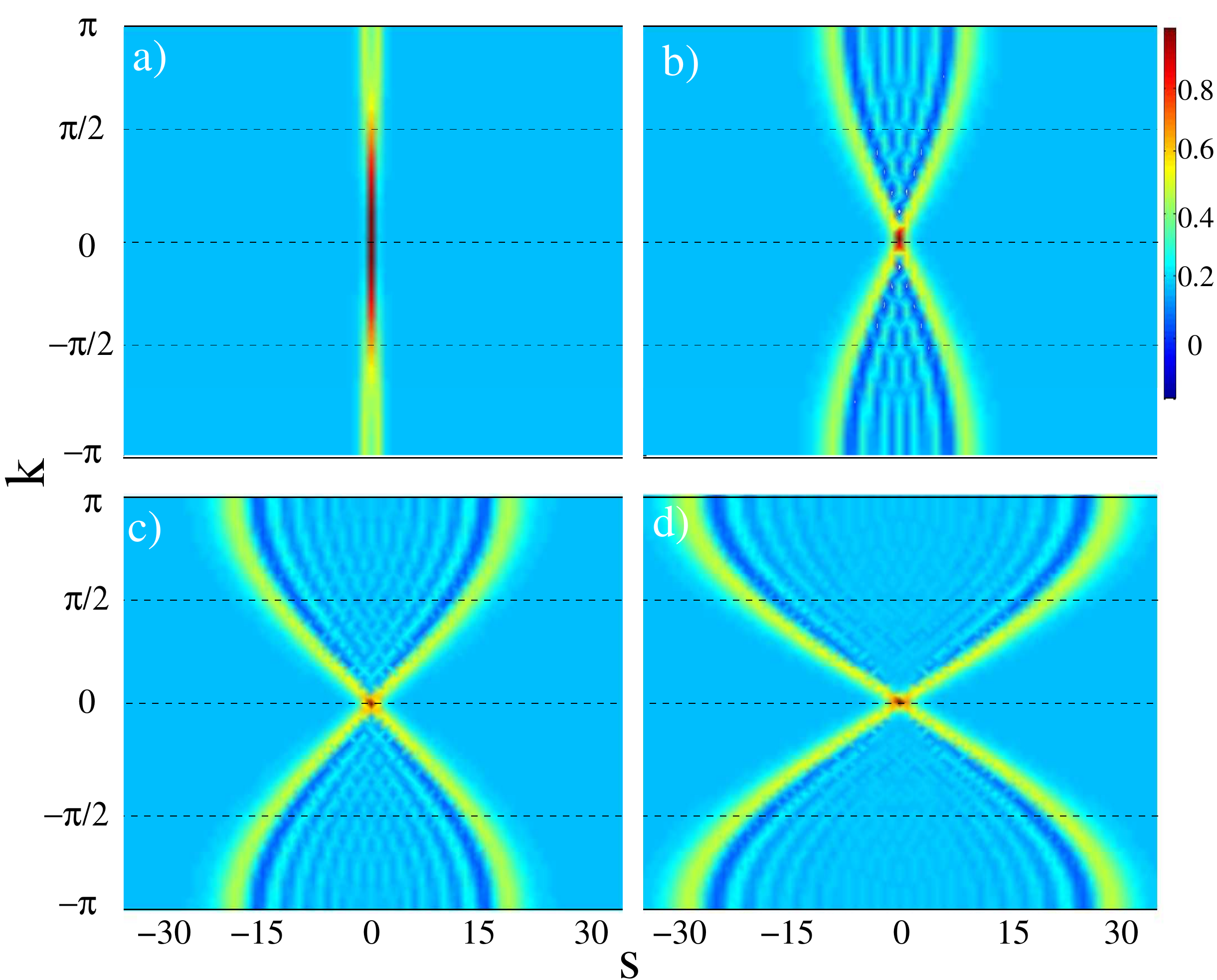}
\caption{ (Color online) Normalized Wigner function $W(s,k,t^{\prime
})/W_{max}$ as a function of position $s$ and momentum $k$, for $t^{\prime
}=1$, $10$, $20$, $30$ ([(a) $-$ (d)]) and $r_{D}=0.05$, in this case $%
W_{max}$ is the maximum value of Wigner function for each value of $%
t^{\prime }$. }
\label{fig-wig2}
\end{figure}

In a complementary way, we now show the Wigner function (similar to Fig \ref%
{fig-wig}), but we have fixed the parameter $r_{D}=0.05$ and changed the
dimensionless time $t^{\prime }$. In Fig.\ref{fig-wig2}[(a)-(d)], we observe
the Wigner function for $t^{\prime }=1$, $10$, $20$, $30$. The
time-evolution of the Wigner function also shows the transition from the DQW
to the CRW.

\begin{figure}[t]
\includegraphics[height=6cm,width=8.5cm]{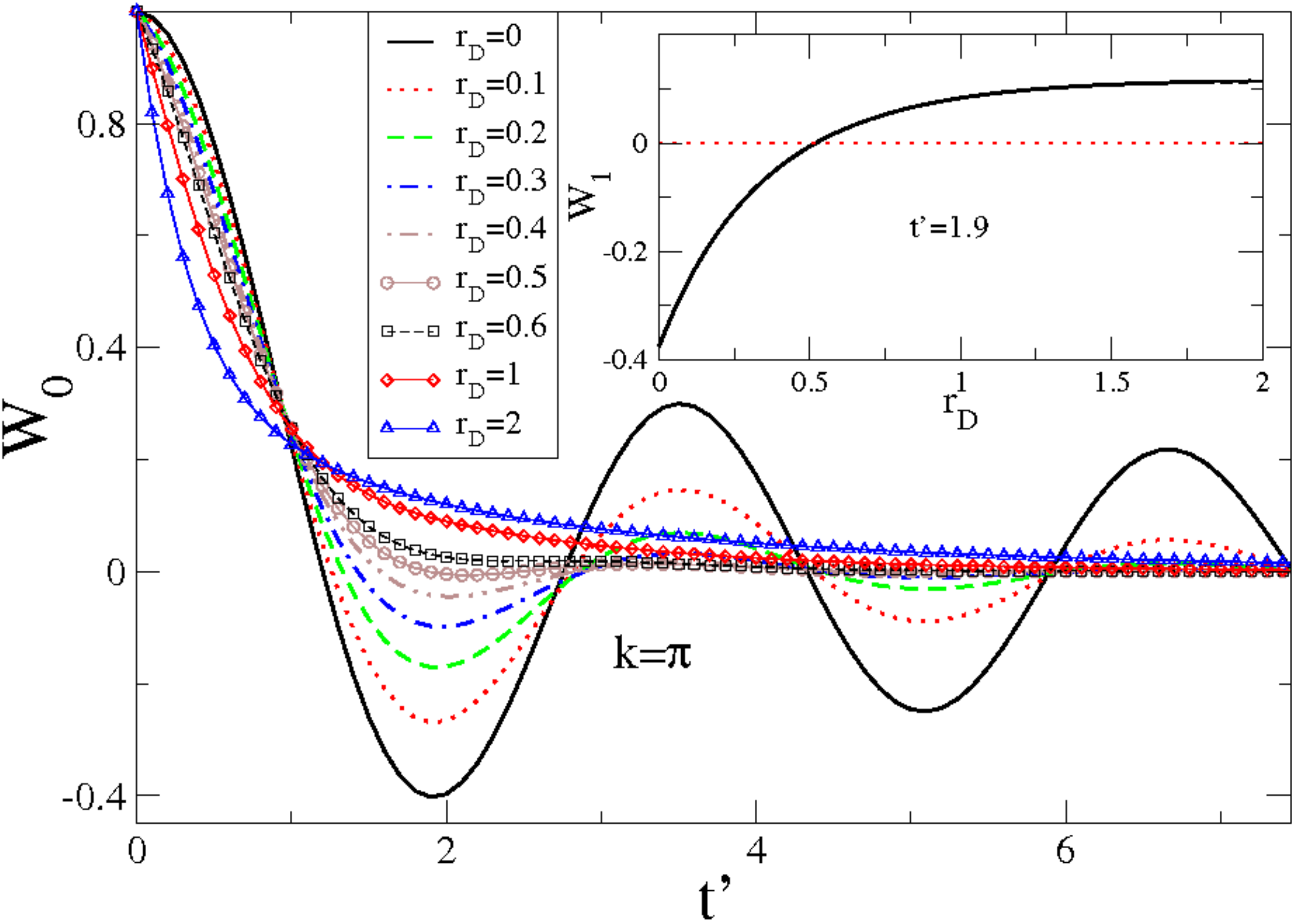}
\caption{ The Wigner function $W_{0}=W(0,\protect\pi ,t^{\prime })$, as a
function of time $t^{\prime }$, for $r_{D}=0$, $0.1$, $0.2$, $0.3$, $0.4$, $%
0.5$, $0.6$, $1$, $2$. The Wigner function $W_{1}=W(0,\protect\pi ,1.9)$, as
a function of $r_{D}$, see inset. For $r_{D}\gtrsim 0.52$ the Wigner
function is non-negative, and well defined as a joint probability density in
the phase space.}
\label{fig-wig3}
\end{figure}

In order to define a criterion to indicate when quantum correlations
dominate over the classical correlations, or vice versa, we show in Fig.\ref%
{fig-wig3} the Wigner function $W_{0}=W(0,\pi ,t^{\prime })$, 
as function of time $t^{\prime }$. We note an oscillatory behavior of $W_{0}$
for $r_{D}<1$. In the inset of Fig.\ref{fig-wig3}, we show the Wigner
function $W_{1}=W(0,\pi ,1.9)$ as a function of $r_{D}$ (we consider the
value of $t^{\prime }$ where $W_{0}$ takes their first minimum). We use this
as a criterion to indicate when quantum correlations are more important than
classical correlations, and from Fig.\ref{fig-wig3}, we note that for $%
r_{D}\lesssim r_{D}^{c}$, with $r_{D}^{c}=0.52$, the Wigner function is $%
W_{0}<0$, therefore, the quantum correlations dominate in the system. For
the case when $r_{D}>r_{D}^{c}$, the Wigner function is non-negative for all
parameters in the system. In this case the classical correlations dominate
over the quantum correlations. Therefore, in this situation, we could say
that the Wigner function is a joint probability density in the phase space
(because $W(s,k,t)\geq 0$). This criterion is not unique, other criteria
could be used.

These results are consistent with the previous results obtained with the
probability of finding the particle in the lattice site (see Figures \ref%
{fig-prob2D}-\ref{fig-prob3D}).

\subsection{The long-time limit of the reduced density matrix}

In this subsection, we have analyzed the long-time limit behavior of the
reduced density matrix. The Eq.(\ref{roL1L2}) can be written in the form:

\begin{equation}
\left\langle s_{1}\right\vert \rho (t)\left\vert s_{2}\right\rangle
=\int_{-\pi }^{\pi }\frac{dk_{1}}{2\pi }e^{i\frac{\Omega t}{\hbar }\cos
k_{1}}\mathcal{I}(k_{1}) ,  \label{LONGtime}
\end{equation}%
where

\begin{equation}
\mathcal{I}(k_{1})\!\!=\!\!\!\int_{-\pi }^{\pi }\!\!\frac{dk_{2}}{2\pi }e^{-i%
\frac{\Omega t}{\hbar }\!\cos k_{2}}\!\cos \left[ \left(
k_{1}\!\!-\!\!k_{2}\right) \!(s_{1}\!\!-\!\!s_{2})\right] e^{2Dt\left( \cos
(k_{1}\!-\!k_{2})\!-\!1\right) },  \label{LT2}
\end{equation}%
By symmetry the term proportional to $\sin \left[ \left( k_{1}-k_{2}\right)
(s_{1}-s_{2})\right] $ cancels out. Noting that in the long-time limit $%
\mathcal{I}(k_{1})$ can be calculated by using the stationary phase
approximation \cite{MSP}, we get (in the limit: $\Omega /\hbar >>2D$ and $%
t\rightarrow \infty $)%
\begin{eqnarray*}
\mathcal{I}(k_{1}) &\simeq &\sqrt{\frac{2\pi }{\Omega t}}\left\{
e^{-2Dt(1-\cos k_{1})-i(\Omega t-\pi /4)}\cos k_{1}s_{1}\right. \\
&&\left. +e^{-2Dt(1+\cos k_{1})+i(\Omega t-\pi /4)}\cos (k_{1}s_{1}-\pi
s_{2})\right\} .
\end{eqnarray*}%
Introducing this expression in Eq.(\ref{LONGtime}) we can apply once again
the stationary phase approximation (taking care that the saddle point $(0,0)$
in the bi-dimensional integration does not contribute), then for the reduced
density matrix in the asymptotic limit ($(\Omega /\hbar )t\rightarrow \infty 
$) we get:

\begin{eqnarray}
\langle s_{1}|\rho (t)|s_{2}\rangle \! &\rightarrow &\!\frac{2\hbar }{\pi
\Omega t}\{\cos \pi (s_{1}+s_{2})+\cos \pi (s_{1}-s_{2})  \nonumber
\label{ro-LT} \\
&&+e^{-4Dt}[\sin (2\frac{\Omega }{\hbar }t)\cos (s_{1}+s_{2})\frac{\pi }{2}%
\cos (s_{1}-s_{2})\frac{\pi }{2}]  \nonumber \\
&&-ie^{-4Dt}[\cos (2\frac{\Omega }{\hbar }t)\sin (s_{1}+s_{2})\frac{\pi }{2}%
\sin (s_{1}-s_{2})\frac{\pi }{2}]\}.  \nonumber \\
&&
\end{eqnarray}%
This result is in agreement with the asymptotic approximation of the Bessel
function when it is replaced in the Eq.(\ref{rhog}) (for $x\rightarrow
\infty $, $J_{n}(x)\approx \sqrt{\frac{2}{\pi x}}\cos (x-\frac{\pi }{4}-n%
\frac{\pi }{2})$). Thus we see that for long-time, terms proportional to $%
e^{-4Dt}$ will contribute to the quantum entropy production associated with
the reduced density matrix. In fact, this contribution is proportional to
the following matrix (in Wannier representation):%
\begin{equation}
\begin{array}{cc}
& \!\!\!\!\!\!\!\!\cdots \cdots \cdots -1,\;\;\;0,\;\;+1\cdots \cdots \cdots
\cdots \\ 
\begin{array}{c}
\vdots \\ 
\vdots \\ 
\vdots \\ 
-1 \\ 
\,\,\,\,0 \\ 
+1 \\ 
\vdots \\ 
\vdots \\ 
\vdots%
\end{array}
& \left( 
\begin{array}{cccccccccc}
\ddots & \cdot & \cdot & \cdot & \cdot & \cdot & \cdot & \cdot & \cdot & 
\cdot \\ 
\cdot & A & -iB & A & -iB & A & -iB & A & \cdot &  \\ 
\cdot & iB & A & iB & A & iB & A & iB & \cdot &  \\ 
\cdot & A & -iB & A & -iB & A & -iB & A & \cdot &  \\ 
\cdot & iB & A & iB & A & iB & A & iB & \cdot &  \\ 
\cdot & A & -iB & A & -iB & A & -iB & A & \cdot &  \\ 
\cdot & iB & A & iB & A & iB & A & iB & \cdot &  \\ 
\cdot & A & -iB & A & -iB & A & -iB & A & \cdot &  \\ 
\cdot & \cdot & \cdot & \cdot & \cdot & \cdot & \cdot & \cdot & \cdot & 
\ddots%
\end{array}%
\right) ,%
\end{array}
\label{matrix}
\end{equation}%
where $A=\frac{2\hbar }{\pi \Omega t}(1-\sin (2\Omega t/\hbar )e^{-4Dt})$, $%
B=\frac{2\hbar }{\pi \Omega t}\cos (2\Omega t/\hbar )e^{-4Dt}$. This means
that asymptotically the probability profile $\rho _{ss}(t)$\ goes to zero
uniformly in the lattice, similar to $\sim \frac{2\hbar }{\pi \Omega t}%
(1-\sin (2\Omega t/\hbar )e^{-4Dt}$, and the off-diagonal elements form a
time-dependent coherent binary structure of $\left\{ A,\pm iB\right\} $ (in
the Wannier representation) that also goes to zero asymptotically as $\sim
1/t$. Due to the fact that Anderson's boundaries move at finite velocity $%
V_{A}=\frac{1}{\sqrt{2}}\frac{\Omega }{\hbar }$ away from the initial
condition (state $\left\vert s_{0}\right\rangle \left\langle
s_{0}\right\vert $), we expect that the DQW will be always inside a
(time-dependent) finite domain of maximum size $L\sim \sqrt{\left\langle 
\mathbf{q}(t)^{2}\right\rangle -\left\langle \mathbf{q}(t)\right\rangle ^{2}}%
>>\epsilon $ ($\epsilon $ is the lattice parameter, we take $\epsilon =1$),
which increases linearly in time. Then, we can calculate the asymptotic
eigenvalues of $\left\langle s_{1}\right\vert \rho (t)\left\vert
s_{2}\right\rangle $ by approximating $\rho (t)$ to be a $L\times L$
finite-domain matrix.

It is simple to calculate the non-null eigenvalues of a matrix of dimension $%
L\times L$ of the form (\ref{matrix}). For $D=0$ (without dissipation) there
is only one non-null eigenvalue:%
\begin{equation}
\lambda =\frac{2\hbar }{\pi \Omega t}[L-\sin (2\frac{\Omega }{\hbar }t)].
\label{eigenD0}
\end{equation}%
On the other hand, for the case $D\neq 0$, but $\Omega /\hbar >>2D$, we
obtain only two non-null eigenvalues for the reduced density matrix. These
eigenvalues are as follows: 
\begin{eqnarray}
\lambda _{\pm } &=&\frac{2\hbar }{\pi \Omega t}\{L-\sin (2\frac{\Omega }{%
\hbar }t)  \nonumber  \label{eigenDne0} \\
&&\pm \sqrt{1\!+\!(L^{2}\!-\!1)e^{-8Dt}\!-\!2Le^{-4Dt}\sin (2\frac{\Omega }{%
\hbar }t)\!+\!e^{-8Dt}\sin ^{2}(2\frac{\Omega }{\hbar }t)}\}.  \nonumber \\
&&
\end{eqnarray}%
%
%
%
%
%
%
%
%
%
%
%
%
%
%
%
%
%
%
%
%
%
%
%
%
%
%
As expected, we observe from this expression that if we take $D=0$, we
recover the Eq.(\ref{eigenD0}). 
These results allow us to calculate the long-time behavior of the von
Neumann entropy.

\section{Quantum entanglement in the DQW}

\subsection{Time evolution of von Neumann's Entropy}

In order to study the irreversibility behavior of a free particle in
interaction with a quantum thermal bath $\mathcal{B}$, i.e., our DQW model,
we have calculated von Neumann's entropy corresponding to non-equilibrium
situations, assuming that the system is prepared in a highly localized state
(i.e., we adopt the initial condition $\rho (0)=\left\vert
s_{0}\right\rangle \left\langle s_{0}\right\vert $, with $s_{0}=0$).

\begin{equation}
\mathcal{S}(t)=-\mbox{Tr}\left[ \rho (t)\ln \rho (t)\right].  \label{entrop}
\end{equation}

\begin{figure}[t]
\caption{(Color online) von Neumann's entropy as a function of $t^{\prime }$
and $r_{D}$ (Fig (a)). This function measures the quantum entanglement
between the particle in the lattice and the phonons bath. We show a plot in
2D (Fig. (b)) of quantum entropy as a function of $t^{\prime }$ for
different values of $r_{D}=0$, $0.01$, $0.05$, $0.1$, $0.5$, $1$, $2$. }
\label{fig-ent}\includegraphics[height=11.cm,width=9.cm]{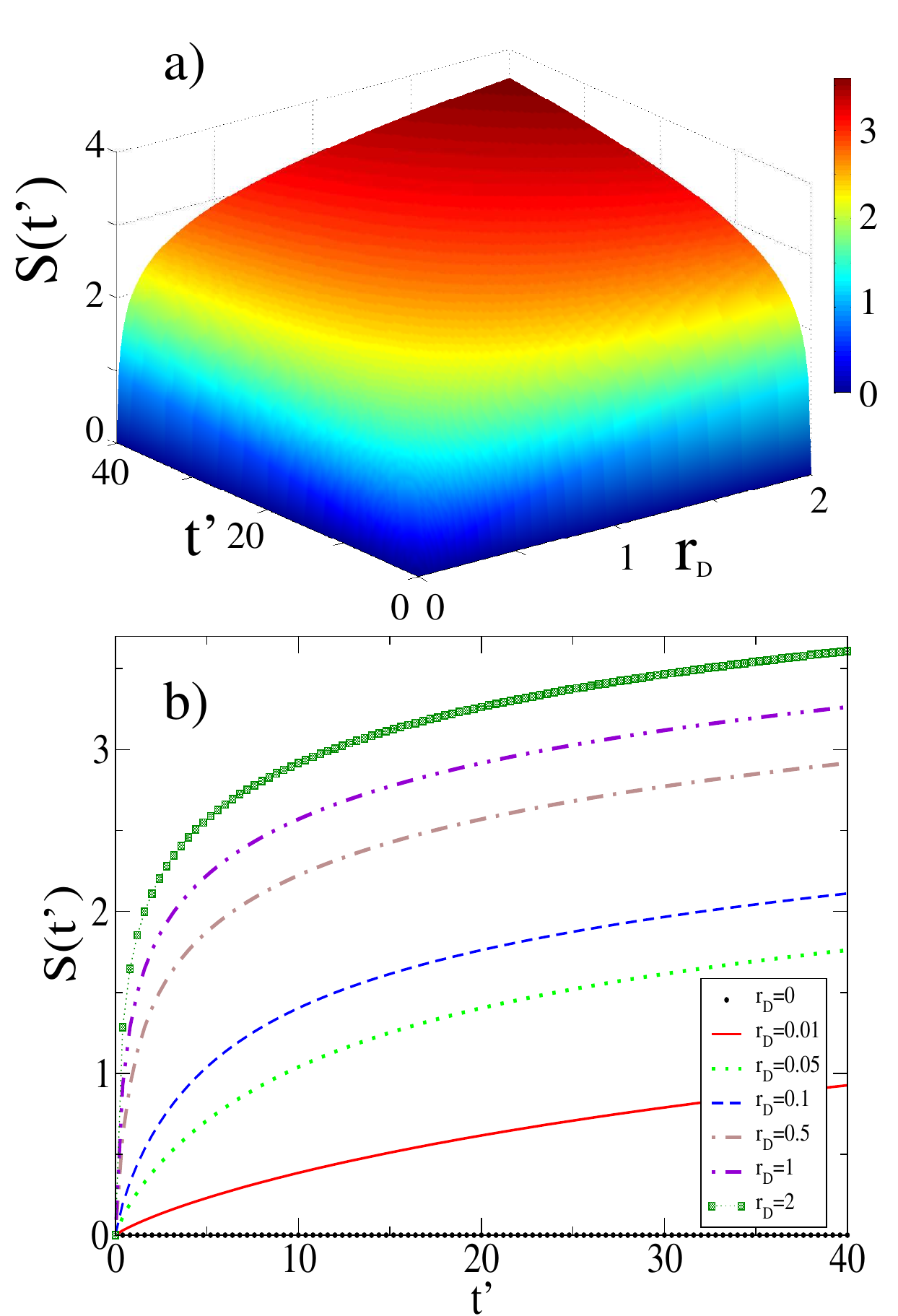}
\end{figure}

In an open quantum system the reduced density matrix evolves in a Markovian
approximation, following the QME (Eq.(\ref{QME})). Due to the dissipation,
the reduced density matrix $\rho (t)$ will not be diagonal at any time $t>0$%
. The information of the quantum entanglement between the quantum thermal
bath and our free particle -in the lattice- can be obtained from the reduced
density matrix given in the Eq.(\ref{rhog}). Due to the fact that the total
system is a pure state, the von Neumann entropy for the reduced density
matrix can be used to measure the entanglement. Then we should calculate von
Neumann's entropy numerically using $\rho (t)$ in the Wannier base. We have
fixed the size of the chain to be $L$, with $L\approx V_{A}t$ where $V_{A}=%
\frac{1}{\sqrt{2}}\frac{\Omega }{\hbar }$, thus we have diagonalized the
reduced density matrix. Then we can use the following expression for the
quantum entropy: 
\begin{equation}
\mathcal{S}(t)=-\sum_{i}\Lambda _{i}\ln \Lambda _{i},  \label{entrop-dia}
\end{equation}%
where $\Lambda _{i}$ is an eigenvalue of the reduced density matrix $\langle
s_{1}\arrowvert\rho (t)\arrowvert s_{2}\rangle $ in the domain $\left[ -L,L%
\right] .$ We expect that for finite times (even in an infinite lattice) $%
\mathcal{S}(t)\neq \infty .$ We have carried out this calculation
numerically and we have shown in Fig.\ref{fig-ent} $\mathcal{S}(t)$ as a
function of time $t^{\prime }=\frac{\Omega }{\hbar }t$ for different values
of the dissipative parameter $r_{D}=\frac{2D}{\Omega /\hbar }$.

In Fig.\ref{fig-ent}(a), we observe the quantum entropy as function of $%
r_{D} $ and $t^{\prime }$ (3D visualization). As expected for $r_{D}=0$ ($%
D=0 $), the quantum entropy is $\mathcal{S}(t^{\prime })=0$ for all $%
t^{\prime }\geq 0$. In this case the quantum entanglement between the free
particle in the lattice and the phonon bath is zero, which means we can
write the wave function of the total system as a product of one state of the
free particle in the lattice and one of the phonon bath (separable state).
Another trivial result is the case that $t^{\prime }=0$, where the quantum
entropy is zero for $r_{D}\geq 0$ (in this case $\rho (t^{\prime }=0)=%
\arrowvert s_{0}\rangle \langle s_{0}\arrowvert$). In the presence of
dissipation the total system is no longer a separable state between the
particle in the lattice and environment. The reduced density matrix $\rho
(t) $ is a mixed state (for $D>0$), and the quantum entropy starts
increasing in time for a fixed value of $r_{D}$. We also note that for a
fixed value of $t^{\prime }$, the magnitude of the quantum entropy increases
as $r_{D}$ increases. The quantum entropy gives information about the
transition from the DQW to the CRW, which is consistent with the results
obtained above concerning the probability profile and the Wigner function.
Finally, in Fig.\ref{fig-ent}(b), we show results of the quantum entropy as
a function of $t^{\prime }$ and different values of $r_{D}$.

Similar analyses to measure quantum correlations have been carried out by
considering a free particle (in a lattice) with an additional internal
degree of freedom: the quantum "coin". Thus, the correlation between the 
\textit{coin} and the \textit{spatial} degree of freedom have been studied
in detail, and these models share some similarities with our results despite
the fact that by considering \textit{the quantum coin} as the thermal bath,
the latter has a finite Hilbert space \cite{chandrashekar,Romanelli2}.

Interestingly, by introducing the stationary phase approximation in Eq.(\ref%
{roL1L2}), it is possible to obtain asymptotic behavior for $\left\langle
s_{2}\right\vert \rho (t)\left\vert s_{1}\right\rangle $, then we can study
the time-dependence of $\mathcal{S}(t)$ in the long-time regime analytically.

\subsubsection{The long-time limit of von Neumann's entropy}

Having the result of Eq.(\ref{ro-LT}) we can re-write von Neumann's entropy
in the long-time regime in the form:%
\begin{eqnarray}
\mathcal{S}(t) &=&-\sum_{s_{1}=-\infty }^{\infty }\sum_{s_{2}=-\infty
}^{\infty }\!\!\!\!\left\langle s_{1}\right\vert \rho (t)\left\vert
s_{2}\right\rangle \left\langle s_{2}\right\vert \ln \rho (t)\left\vert
s_{1}\right\rangle  \nonumber \\
&\simeq &-\sum_{j=1}^{L}\lambda _{j}\ln \lambda _{j},  \label{HLT}
\end{eqnarray}%
here $\lambda _{j}$ are the eigenvalues associated with the asymptotic
long-time regime of the reduced density matrix $\left\langle
s_{1}\right\vert \rho (t)\left\vert s_{2}\right\rangle $. Using Eq.(\ref%
{eigenDne0}) in Eq.(\ref{HLT}) and considering $L>>1$ (for simplicity, we
re-normalized the eigenvalues of $\rho (t)$); we get (when $\frac{\Omega }{%
\hbar }>>2D$ and $\frac{\Omega }{\hbar }t>>1$):%
\begin{eqnarray}
\mathcal{S}(t)\!\! &\simeq &-(\frac{1-e^{-4Dt}}{2})\ln (\frac{1-e^{-4Dt}}{2})
\nonumber \\
&-&(\frac{1+e^{-4Dt}}{2})\ln (\frac{1+e^{-4Dt}}{2}).  \label{eq-ent-t}
\end{eqnarray}

In addition, another asymptotic approximation to $\mathcal{S}(t)$ can be
made if we consider $D\sim 0$ (with $Dt<<1$ ), so we can approximate $%
e^{-4Dt}\rightarrow 1-4Dt$ and replacing this expression in Eq.(\ref%
{eq-ent-t}) we get a simpler expression for the quantum entropy in the form:

\[
\mathcal{S}(t)\simeq -2Dt\ln (2Dt),\quad Dt<<1. 
\]%
Note that in the limit $D\rightarrow 0$ (even when $t\gg 1$) the quantum
entropy $\mathcal{S}(t)\rightarrow 0$. We have checked this analytical
result with our numerical calculation using Eq.(\ref{entrop-dia}) and the
agreement is excellent, see Fig.\ref{fig-ent}.

\section{Conclusions}

A free particle -in an infinite regular lattice- in interaction with a
thermal phonon bath has been studied by tracing out the degree of freedom of
the bath. We then worked out the quantum master equation for a dissipative
tight-binding model.

We have solved the master equation analytically by using Bessel functions,
and obtained the reduced density matrix $\rho (t)$ as a function of the
scale energies of the system ($\Omega /\hbar $, $D$). We have also studied
the transition from Dissipative Quantum Walk to Classical Random Walk in
terms of parameters $D$ and $\Omega /\hbar $ (or $r_{D}=\frac{2D}{\Omega
/\hbar }$). In the case when $2D<<\Omega /\hbar $ ($r_{D}<<1$) the quantum
behavior is more important than the dissipation in the system. In the
opposite case we re-obtain the Classical Random Walk $2D>>\Omega /\hbar $ ($%
r_{D}>>1$) because in this case the dissipation creates decoherence in the
system. We have studied the Wigner function to analyze the
pseudo-probability densities in the phase space. This function is very
useful as an indicator of this quantum-classical transition.

As an alternative approach to the study of the transition from Dissipative
Quantum Walk to Classical Random Walk we have used tools from quantum
information theory (as a function of dissipative parameter) to analyze the
reduced density matrix. To describe this transition we have used von
Neumann's entropy $\mathcal{S}(t)$ to measure the quantum entanglement
between the free particle -in a lattice- and the phonon bath. We observed
that for $D=0$ the quantum entropy is $\mathcal{S}(t^{\prime })=0$ for $%
t^{\prime }\geq 0$ (closed system), and when $D$ increases we show that
quantum decoherence starts to appear and therefore the quantum entropy
increases in time with a law which is slower than that from classical
statistics ($\mathcal{S}(t)_{CRW}\sim \ln t$) \cite{libro,Kampen}.
Asymptotically for $D\rightarrow 0$ the quantum entropy turns out to be only
a function of the dissipative parameter $D$. This fact also indicates the
beginning of the transition from the Dissipative Quantum Walk to the
Classical Random Walk.

This analytical model allows us to study the effect of decoherence in the
Dissipative Quantum Walk as a function of the two typical energies of the
system. We can conclude that in the present model there are two
characteristic time scales: the dissipative time $\tau _{D}\sim 1/D$ and the
hopping time $\tau _{H}\sim \hbar /\Omega $; the competition between these
time scales controls the decoherence and correlation mechanism in the
system. For example, the quantum purity $\mathcal{P}_{Q}(t)$ is controlled
by $\tau _{D}$, but in general the entropy and the interference phenomena
appearing in the probability profile or in the Wigner phase-space
pseudo-distribution are controlled by the competition between these time
scales.

The interesting problem of the propagation of photons in waveguide lattices
are possible scenarios where our present results can be applied, also the
analysis of the entanglement of a bipartite system can be studied in the
present framework, works along these lines are in progress. In this way the
present model gives insight into the effect of dissipation in more complex
quantum systems, for instance, the analysis of quantum correlations between
two particles -in a regular lattice- in interaction with a phonon bath.

\textit{Acknowledgments}. We thank Maria del Carmen Ferreiro for the English
revision of the manuscript. M.O.C gratefully acknowledges support received
for this study from Universidad Nacional de Cuyo, Argentina, project SECTyP,
and CONICET, Argentina, grant PIP 90100290. M.N. gratefully acknowledges
CONICET, Argentina for his Post-Doctoral fellowship.

\appendix

\section{On the second quantization and the one-particle tight-binding
Hamiltonian}

\label{ap1:sec-quan}

A free Hamiltonian in the tight-binding approximation for spinless particles
(fermion) can be written in second quantization in the form \cite{haken}:%
\begin{equation}
H_{S}=E_{0}\sum_{s_{=}{-\infty }}^{\infty }c_{s}^{\dag }c_{s}-\frac{\Omega }{%
2}\left( \sum_{s_{=}{-\infty }}^{\infty }c_{s-1}^{\dag }c_{s}+c_{s+1}^{\dag
}c_{s}\right) ,  \label{Hamiltonian-s-q}
\end{equation}%
where $c_{s}^{\dag }$ and $c_{s}$ are creation and destruction operators in
the site $s$ of the lattice respectively ($\arrowvert\cdots
,0,1_{s},0,\cdots \rangle =c_{s}^{\dag }\arrowvert0\rangle $, where $%
\arrowvert0\rangle $ is the empty state). Then considering only one particle
it is straightforward to compare Eq.(\ref{Hamiltonian-s-q}) with $H_{S}$ in
Eq.(\ref{Hamiltonian}), if we replace $a$ and $a^{\dag }$ with a combination
of $c_{s^{\prime }}^{\dag }$ and $c_{s^{\prime }}$, in the following way:

\begin{equation}
a\Rightarrow R=\sum_{s_{=}{-\infty }}^{\infty }c_{s-1}^{\dag
}c_{s},\;\;\;a^{\dag }\Rightarrow R^{\dag }=\sum_{s_{=}{-\infty }}^{\infty
}c_{s+1}^{\dag }c_{s},  \label{a-R}
\end{equation}%
where $c_{s}$ are acting in the Fock-space. Therefore, we can also check
that $R$ and $R^{\dag }$ commute in the general case for many particles
(where $R^{\dag }R=RR^{\dag }=\sum_{s_{=}{-\infty }}^{\infty }c_{s}^{\dag
}c_{s}-\sum_{s,s_{=}^{\prime }{-\infty }}^{\infty }c_{s+1}^{\dag
}c_{s^{\prime }-1}^{\dag }c_{s}c_{s^{\prime }}$), and for one particle in
the lattice we get $RR^{\dag }=\mathbf{1}$. 

Equation (\ref{a-R}) shows the expected mapping from Fock's space into the
Winner basis. Then the connection between the tight-binding Hamiltonian and
the QW model can be established.

\section{Reduced matrix density}

\label{ap:matrix} Here we show how to obtain Eq.(\ref{rhog}). Replacing the
following relations for the Bessel function: $e^{iz\cos \theta
}=\sum_{n=-\infty }^{\infty }i^{n}J_{n}(z)e^{in\theta }$; $e^{z\cos \theta
}=\sum_{n=-\infty }^{\infty }I_{n}(z)e^{in\theta }$ in Eq.(\ref{roL1L2}),
where $J_{n}$ and $I_{n}$ are a Bessel functions of integer order \cite%
{abramowitz,evangelidis}, we find: 
\begin{eqnarray}
\langle s_{1}|\rho (t)|s_{2}\rangle &\!\!\!\!=\!\!\!\!&\left( \!\frac{1}{%
2\pi }\!\right)
^{\!\!2}e^{-2Dt}\!\!\!\!\!\!\!\!\!\sum_{m_{1},m_{2},n=-\infty }^{\infty
}\!\!\!\!\!\!\!\!J_{m_{1}}\!\!\left( \!\frac{\Omega t}{\hbar }\!\right)
\!\!J_{m_{2}}\!\!\left( \!\frac{\Omega t}{\hbar }\!\right) \!\!I_{n}\!\left(
2Dt\right)  \nonumber \\
&&\times i^{m_{1}+m_{2}}\!\!\int_{-\pi }^{\pi
}\!\!\!\!dk_{1}e^{i(s_{1}+m_{2}+n)k_{1}}\!\!\int_{-\pi }^{\pi
}\!\!\!\!dk_{2}e^{-i(s_{2}-m_{1}+n)k_{2}}.  \nonumber
\end{eqnarray}

Using the definition of Kronecker delta: $\delta _{s,s^{\prime }}=\frac{1}{%
2\pi }\int_{-\pi }^{\pi }e^{ik(s-s^{\prime })}$ in the previous expression,
we finally obtain: 
\[
\langle s_{1}|\rho (t)|s_{2}\rangle
\!\!=\!i^{(s_{1}-s_{2})}e^{-2Dt}\!\!\!\!\!\sum_{n=-\infty }^{\infty
}\!\!\!\!J_{s_{1}+n}\!\!\left( \!\frac{\Omega t}{\hbar }\!\right)
\!\!J_{s_{2}+n}\!\!\left( \!\frac{\Omega t}{\hbar }\!\right)
\!\!I_{n}\!\left( 2Dt\right) . 
\]

\section{Moments of the position operator - The characteristic function}

\label{ap2:momen} We have defined a characteristic function \cite{Kampen}
for calculating moments of position operator $\mathbf{q}$ in the following
way:

\begin{equation}
\mathcal{G}(\xi )=\mbox{Tr}\lbrack \rho (t)e^{i\xi \mbox{{\bf q}}%
}]=\sum_{l=-\infty }^{\infty }\langle l\arrowvert\rho (t)\arrowvert l\rangle
e^{i\xi l},  \label{fgen}
\end{equation}%
thus the quantum moments of $\mathbf{q}$ can be obtained using the following
expression: 
\begin{equation}
\langle \mbox{{\bf q}}(t)^{m}\rangle =\frac{1}{i^{m}}\frac{d^{m}}{d\xi ^{m}}%
\mathcal{G}(\xi )\biggl|_{\xi =0},  \label{momg}
\end{equation}

Using Eq.(\ref{rhog}) in Eq.(\ref{fgen}), we can write the characteristic
function in the form: 
\begin{equation}
\mathcal{G}(\xi )=e^{-2Dt\left( 1-\cos \xi \right) }\ J_{0}\left( 2t\frac{%
\Omega }{\hbar }\sin \frac{\xi }{2}\right) ,  \label{fgen1}
\end{equation}%
here we have said that $\sum_{n=-\infty }^{\infty }e^{in\gamma
}J_{n+m}(x)J_{n}(x)=J_{m}(2x\sin (\gamma /2))e^{i\beta m}$, where $\beta
=\pi /2-\gamma /2$. From this characteristic function all the moments of the
position operator can be calculated straightforwardly. In particular, we can
re-obtain the variance of the DQW (see Eq.(\ref{variance})). Note that in
the classical limit $\Omega =0$ we recover the expected characteristic
function associated with the CRW \cite{Kampen,libro}. In general, Eq. (\ref%
{fgen1}) shows that the non-equilibrium behavior of the characteristic
function of the DQW is the product of the classical one and the quantum
characteristic function $J_{0}\left( 2t\frac{\Omega }{\hbar }\sin \frac{\xi 
}{2}\right) $.

\end{document}